**Vincent ROCHER**[1*], **Cédric JOIN**[2,6], **Stéphane MOTTELET**[3], **Jean BERNIER**[1], **Sabrina RECHDAOUI-GUERIN**[1], **Sam AZIMI**[1], **Paul LESSARD**[4], **André PAUSS**[3], **Michel FLIESS**[5,6]

[1]SIAAP (Syndicat Interdépartemental pour l'Assainissement de l'Agglomération Parisienne), Direction du Développement et de la Prospective, 82 avenue Kléber, 92700 Colombes, France
[2]CRAN (CNRS, UMR 7039), Université de Lorraine, BP 239, 54506 Vandœuvre-lès-Nancy, France
[3]TIMR (EA 4297), Sorbonne Universités & Université de Technologie de Compiègne, rue du docteur Schweitzer, 60203 Compiègne, France
[4]Département de génie civil et de génie des eaux, Université Laval, Québec, Canada, G1V 0A6
[5]LIX (CNRS, UMR 7161), Ecole polytechnique, 91128 Palaiseau, France
[6]AL.I.E.N. (ALgèbre pour Identification & Estimation Numériques), 24-30 rue Lionnois, BP 60120, 54003 Nancy, France
*vincent.rocher@siaap.fr

**Titre : La production de nitrites lors de la dénitrification des eaux usées par biofiltration - Stratégie de contrôle et de réduction des concentrations résiduelles**

**Title : Nitrite production during wastewater denitrification by biofiltration - a control strategy towards the decrease of residual concentrations**

Titre court : Contrôle des nitrites en dénitrification sur biofiltre

**Résumé**

Le développement des unités de post-dénitrification dans les stations d'épuration de l'agglomération parisienne a fait ré-émerger la problématique du nitrite dans les eaux de Seine en aval de Paris. Le contrôle de l'apparition des nitrites au cours de l'étape de post-dénitrification est donc devenu un enjeu technique majeur. Des études visant à mieux appréhender les mécanismes d'apparition du nitrite lors de la dénitrification des eaux usées et à étudier des évolutions techniques (métrologie et boucles de contrôle-commande des procédés) à mettre en œuvre sur les usines pour contrôler et limiter sa production ont été engagées dans le cadre du programme MOCOPEE (www.mocopee.com). De précédents travaux ont montré que les modes usuels d'injection du méthanol ne permettent pas de s'assurer de la stabilité du rapport C/N dans le réacteur biologique et conduisent à une production erratique et incontrôlée de nitrites. La possibilité d'ajouter une « commande sans modèle » à la commande classique a donc été testée à l'aide du modèle mathématique SimBio, modèle permettant de simuler le fonctionnement des unités de biofiltration. La commande sans modèle placée « en fin de traitement » et basée sur la concentration en nitrites mesurée en sortie de procédé, se greffe à la méthode de contrôle classique en y apportant des corrections seulement au besoin. Les résultats des simulations montrent qu'une régulation des injections de méthanol basée sur la « commande sans modèle » permet de stabiliser et maitriser le nitrite dans le rejet, sans induire d'augmentation des quantités de méthanol injectées.

## Abstract

The recent popularity of post-denitrification processes in the greater Paris area wastewater treatment plants has caused a resurgence of the presence of nitrite in the Seine River. Controlling the production of nitrite during the post-denitrification process has thus become a major technical issue. Research studies have been led in the MOCOPEE program (www.mocopee.com) to better understand the underlying mechanisms behind the production of nitrite during wastewater denitrification and to develop technical tools (measurement and control solutions) to assist on-site reductions of nitrite production. Prior studies have shown that typical methanol dosage strategies produce a varying carbon-to-nitrogen ratio in the reactor, which in turn leads to unstable nitrite concentrations in the effluent. The possibility of adding a model-free control to the actual classical dosage strategy has thus been tested on the SimBio model, which simulates the behavior of wastewater biofilters. The corresponding "intelligent" feedback loop, which is using effluent nitrite concentrations, compensates the classical strategy only when needed. Simulation results show a clear improvement in average nitrite concentration level and level stability in the effluent, without a notable overcost in methanol.

## Mots-cles

## Keywords

# 1. LA PROBLEMATIQUE DU « NITRITE » EN AGGLOMERATION PARISIENNE

## 1.1 Evolution du contexte réglementaire encadrant le traitement des ERU

Ces 20 dernières années, la réglementation concernant le traitement des eaux résiduaires urbaines (ERU) et la qualité des eaux rejetées dans le milieu naturel a fortement évolué. En 2000, la Directive Cadre sur l'Eau (décembre 2000) a notamment imposé aux états membres de l'Union Européenne de restaurer le bon état écologique et chimique des masses d'eau superficielle dans un délai de 15 ans. Le bon état physico-chimique a été défini et des concentrations à ne pas dépasser dans les eaux superficielles ont été fixées pour les paramètres physiques (oxygène, température) et les nutriments (carbone, azote, phosphore). Dans le cas des composés azotés, les seuils à ne pas dépasser ont été fixés à 0,5 mg/L, 0,3 mg/L et 50 mg/L respectivement pour l'azote ammoniacal ($NH_4^+$), l'azote nitreux ($NO_2^-$) et l'azote nitrique ($NO_3^-$).

## 1.2 Evolution de l'outil industriel de traitement des ERU en agglomération parisienne

L'accroissement des exigences réglementaires a conduit les collectivités en charge du traitement des eaux résiduaires urbaines à moderniser leurs usines afin de disposer de filières de traitement capables d'éliminer de manière efficace le carbone, l'azote et le phosphore des eaux usées. En particulier, des efforts importants ont été consentis afin d'intégrer dans les filières de traitement des eaux des unités de traitement biologique de l'azote permettant l'oxydation de l'azote ammoniacal en azote nitrique (étape de nitrification) puis sa réduction en azote gazeux (étape de dénitrification). Ainsi, en région parisienne, le SIAAP traite aujourd'hui l'azote sur l'ensemble de ses usines. Les 2,5 millions de mètres cubes d'eau usée générés sur le bassin versant du SIAAP sont acheminés vers des usines d'épuration qui disposent de filières de traitement biologique permettant la nitrification et la dénitrification des eaux usées (figure 1).

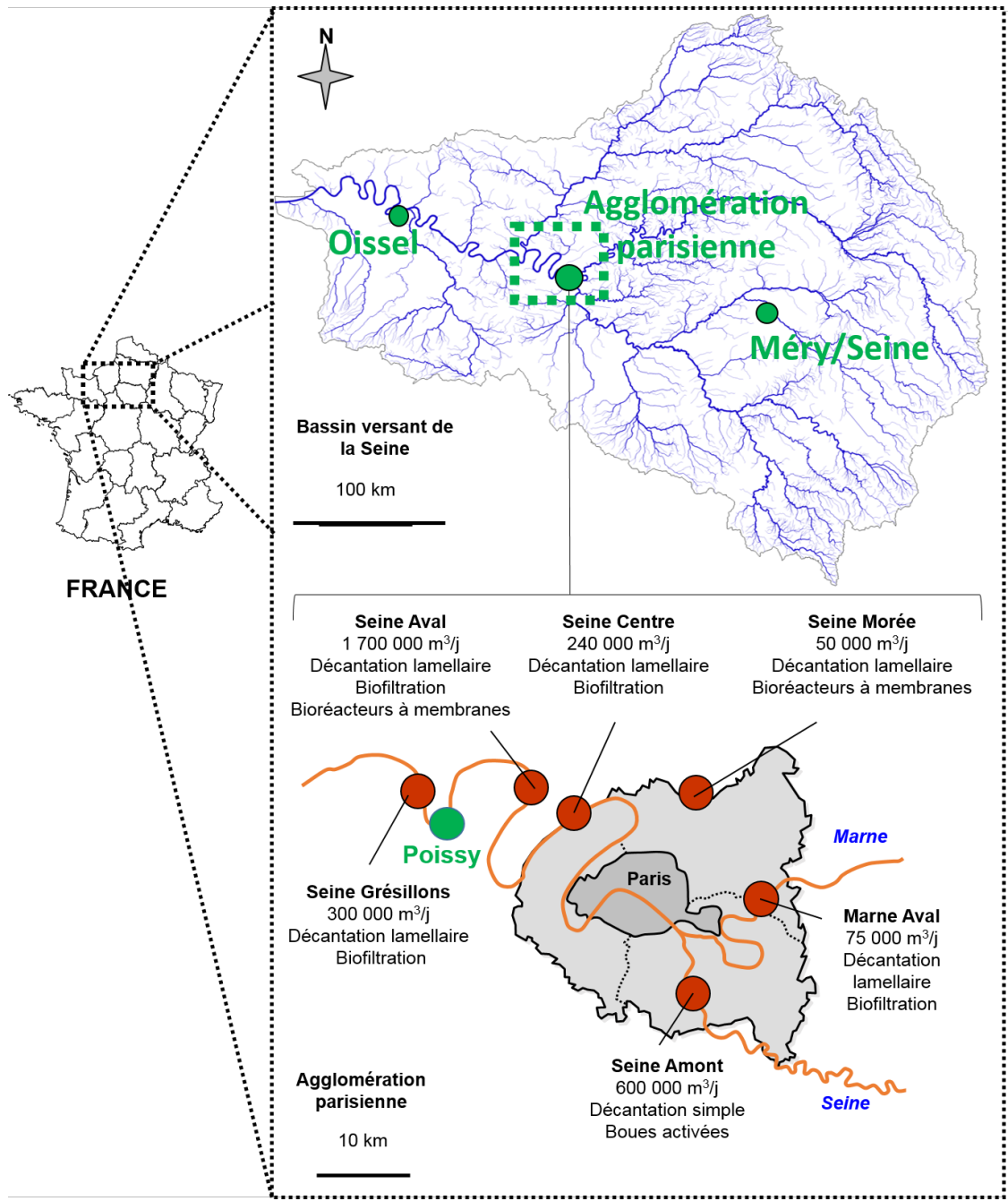

**Figure 1. Bassin versant de la Seine et localisation et caractéristiques des stations d'épuration traitant les eaux usées de l'agglomération parisienne**

Parmi les technologies disponibles pour traiter biologiquement l'azote, la technologie « biofiltration », très compacte, a été largement déployée dans les stations d'épuration parisienne. Dans le cas d'une filière « biofiltration », le traitement de l'azote se fait généralement en associant successivement les étapes de pré-dénitrification, nitrification et post-dénitrification sur méthanol. Dans cette configuration type, les nitrates formés lors de la nitrification sont quasiment totalement éliminés lors de la dénitrification ; les étapes de pré et de post-dénitrification éliminent environ chacune la moitié des nitrates formés (ROCHER et al., 2012 a,b). Dans le cas des installations parisiennes, sur les 2,1 millions de m$^3$ d'eau usée nitrifiés quotidiennement sur biofiltres, environ 1 300 000 m$^3$ sont à ce jour dénitrifiés sur des biofiltres post-dénitrifiants. Il convient néanmoins de souligner que cette proportion sera diminuée lorsque les travaux de modernisation de l'usine Seine Aval (Projet BioSav – Horizon 2022), qui vont conduire à une modification forte de la file de traitement biologique (intégration d'une filière membranaire, notamment), auront été achevés.

**1.3 Emergence récente de la problématique du « nitrite » en Seine**

La politique d'aménagement et de modernisation des filières de traitement des eaux engagée ces dernières années a permis d'améliorer significativement la qualité de la Seine, notamment vis-à-vis des paramètres azotés. Cette amélioration est illustrée par la figure 2a qui présente l'évolution des concentrations en ammonium mesurées ces 10 dernières années au niveau du site de Poissy. Ce point géographique a été sélectionné dans la mesure où il intègre les principaux rejets du SIAAP et, en particulier, les rejets des stations Seine Centre et Seine Aval. On note que, suite à la mise en service de l'unité de nitrification de la station Seine Aval (SAV), les concentrations en azote ammoniacal dans le milieu naturel ont significativement diminué. En 2005, les concentrations en $NH_4^+$ étaient de l'ordre de 2-5 mg/L et sont quasi-systématiquement inférieures à 0,5-1 mg/L depuis début 2007. Les pics de concentrations encore observés de manière erratique résultent des rejets urbains de temps de pluie susceptibles d'introduire ponctuellement dans le milieu des flux importants d'azote ammoniacal ou des périodes de chômage des ouvrages de traitement. D'un point de vue strictement réglementaire, cette diminution des concentrations en azote ammoniacal dans le milieu s'est traduite par une amélioration de la qualité de cette masse d'eau, de l'état « médiocre » à l'état « passable ».

Si le développement des unités de traitement de l'azote a significativement amélioré la qualité des eaux de surface vis-à-vis de l'azote ammoniacal, elle a conjointement fait émerger la problématique du nitrite. La figure 2b, qui présente l'évolution des concentrations en nitrites mesurées ces 10 dernières années au niveau du site de Poissy, montre en effet que la mise en place de l'unité de post-dénitrification sur le site Seine Aval a engendré une augmentation des concentrations en nitrites en Seine. Avant 2007, les niveaux de concentration en nitrites étaient de l'ordre de 0,3-0,4 mg/L. La présence de ces nitrites résultait principalement du processus de nitrification qui s'opérait dans le fleuve et, dans une moindre mesure, des apports des stations d'épuration déjà équipées de post-dénitrification (Seine Centre, Colombes). Depuis 2007, année de la mise en place de l'unité de post-dénitrification sur Seine Aval, les concentrations en nitrites fluctuent entre 0,2 et 1 mg/L ; la limite du bon état étant fixée à 0,3 mg/L.

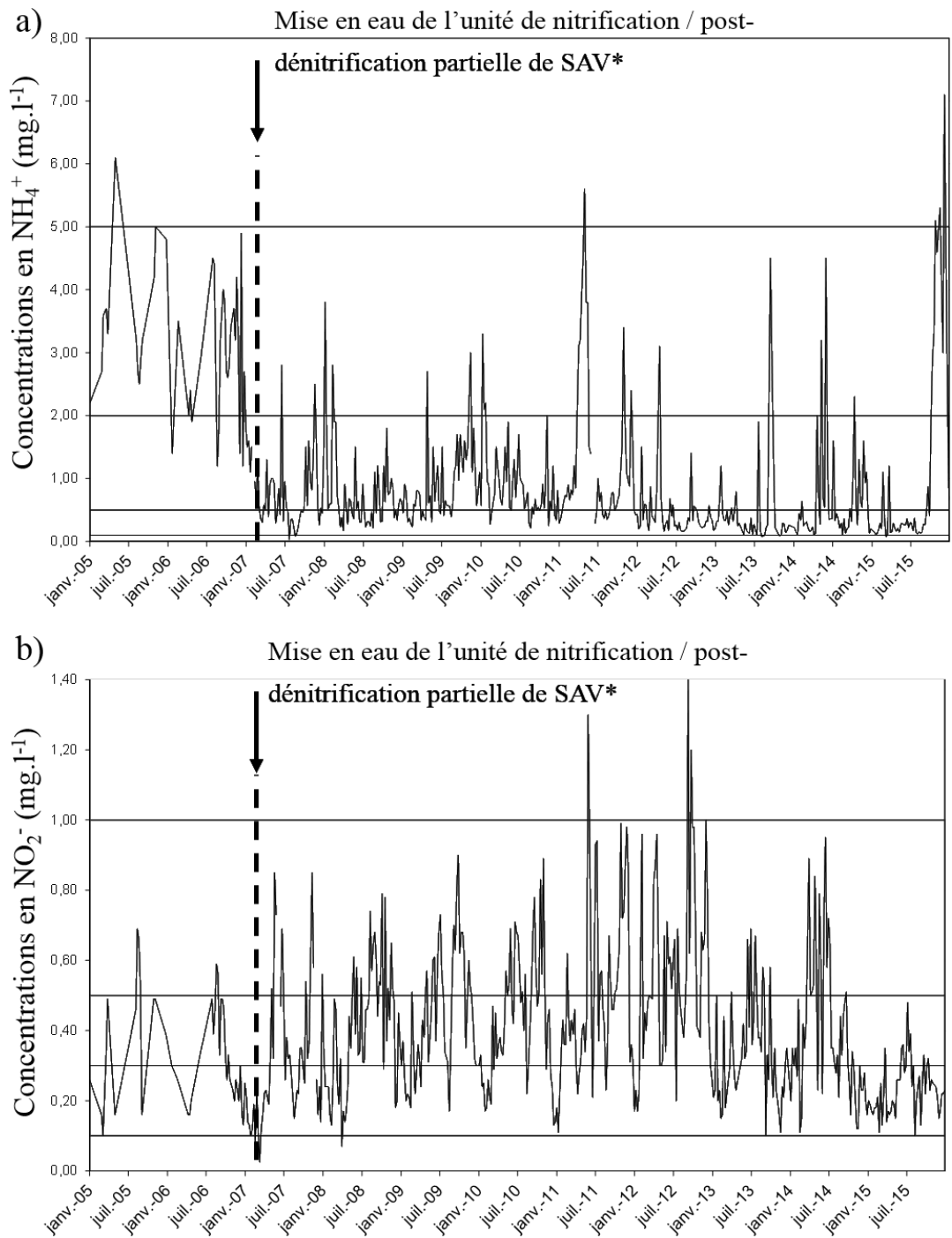

**Figure 2. Evolution de la qualité de la Seine vis-à-vis des paramètres $NH_4^+$ et $NO_2^-$ en aval de l'agglomération parisienne (site de Poissy) entre 2005 et 2015 (Données du réseau de mesure MeSeine). Les lignes horizontales illustrent les limites supérieures du très bon état (0,1**

**mgNH$_4$/L – 0,1 mgNO$_2$/L), du bon état (0,5 mgNH$_4$/L – 0,3 mgNO$_2$/L), de l'état moyen (2 mgNH$_4$/L – 0,5 mgNO$_2$/L) et de l'état médiocre (5 mgNH$_4$/L –1 mgNO$_2$/L) selon la DCE**

Cette augmentation des nitrites en Seine observée à l'aval des stations du SIAAP ne constitue pas une problématique locale puisque les concentrations en nitrites dans les eaux de Seine restent élevées plus de 100 km après le rejet de SAV. Le caractère régional de la problématique « nitrites » est illustré par la figure 3 qui présente l'évolution des concentrations en nitrites en Seine de Méry, 210 km en l'amont de Paris, jusqu'à Oissel, site localisé à plus de 150 km de l'usine de Seine Aval. On note que le pic de concentrations en nitrites observé après les principaux apports urbains (Poissy / Triel) s'estompe très lentement et les concentrations en nitrites mesurées à Poses et Oissel, respectivement à 125 et 154 km de l'usine SAV, sont encore de l'ordre de 0,1-0,4 mg/L.

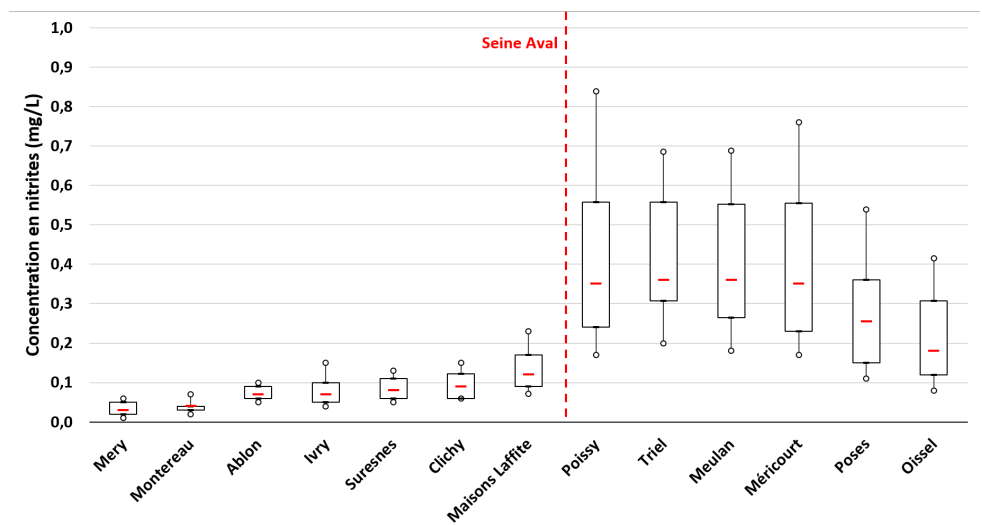

**Figure 3. Evolution des concentrations en nitrites en Seine de Mery à Oissel (Base de données 2010-2015 [SNS], n = 36-262). La boîte est construite à partir des quartiles et médianes (Q25 / médiane / Q75), les extrémités des moustaches représentent les déciles 10 et 90**

Enfin, il convient de rappeler que la présence de nitrites dans les eaux de surface à l'aval de l'agglomération parisienne n'est pas un phénomène récent. Ainsi, dès les années 1970, les concentrations en nitrites dans la rivière étaient élevées (figure 4). Des concentrations de 1,90 mg/L (centile 90) étaient atteintes sur les tronçons de Seine allant de Conflans Sainte-Honorine à Meulan. Dans les années 1994-1995, on observait encore des concentrations importantes en nitrites, de l'ordre de 1,4 mg/L (centile 90), plus à l'aval, de Bonnières à Oissel. Il est probable que ces nitrites, mesurés dans les années 1970-1990, étaient en grande partie produits lors de la nitrification qui s'opérait dans la rivière du fait de la présence d'azote ammoniacal, les stations d'épuration de l'époque ne réalisant pas ou peu la nitrification.

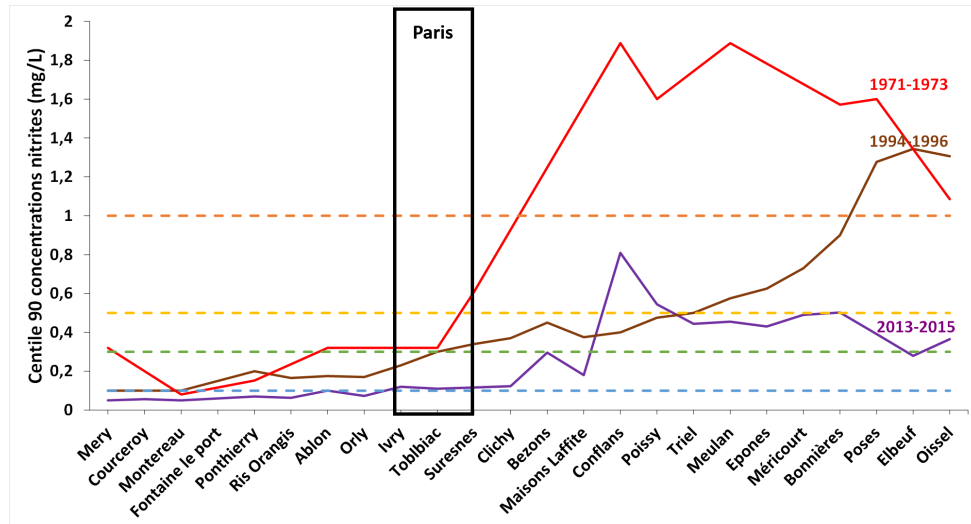

**Figure 4. Evolution des centiles 90 des concentrations en nitrites en Seine de Mery à Oissel pour les années 1971-1973, 1994-1996 et 2013-2015 (Base de données 1970-2015 [SNS], n = 12-132). Les lignes horizontales pointillées illustrent les limites supérieures du très bon état (0,1 mgNO$_2$/L), du bon état (0,3 mgNO$_2$/L), de l'état moyen (0,5 mgNO$_2$/L) et de l'état médiocre (1 mgNO$_2$/L) selon la DCE**

## 2. LIMITES DES MODES D'INJECTION DU METHANOL ACTUELS EN POST-DENITRIFICATION SUR BIOFILTRES

Des études visant à mieux appréhender les mécanismes d'apparition des nitrites lors de la dénitrification des eaux usées et ainsi être capable d'apporter des solutions concrètes pour réduire, voire supprimer, leur production ont récemment été menées (ROCHER et al., 2011a,b,c ; ROCHER et al, 2015). Ces études ont montré que lorsque que les charges en azote appliquées sur les biofiltres dénitrifiant ne dépassent pas la charge admissible et que les concentrations en ortho-phosphates dans l'influent ne sont pas limitantes, le facteur clé qui conditionne l'apparition de nitrites est la régulation des injections de carbone, et notamment de méthanol.

Le ratio DBO injectée / N-NOx appliqué constitue un facteur de contrôle déterminant de l'apparition de nitrites. Le maintien d'un rapport d'environ C/N de 3 dans le réacteur biologique permet d'éviter la présence de nitrites dans les eaux de rejet. Il faut donc que l'exploitant soit capable de maintenir, en toutes circonstances, ce ratio dans les réacteurs biologiques, et cela malgré la fluctuation des conditions d'alimentation des ouvrages dénitrifiants (variations des débits et des concentrations). Cette capacité à stabiliser le rapport C/N dans le réacteur biologique dépend essentiellement du mode de régulation des injections de méthanol appliqué sur l'installation.

De manière schématique, les modes actuels de régulation du méthanol consistent d'abord à évaluer, en temps réel, le flux d'azote accepté sur les unités de dénitrification. Ce flux est généralement estimé à partir de mesures en continu des débits et des concentrations en azote dans les eaux acceptées sur l'ouvrage. Puis, le flux d'azote à éliminer est généralement déterminé en fixant une valeur de consigne en nitrates à respecter dans les eaux de rejet. La quantité de méthanol à injecter est alors calculée en multipliant le flux d'azote à éliminer par un coefficient d'exploitation (K). Ce coefficient d'exploitation, généralement de l'ordre de 3, correspond au rapport entre la quantité de matière organique biodégradable consommée et la quantité de nitrates éliminée au cours de la dénitrification (figure 5). Ce coefficient est fixé à partir de la connaissance métier du système exploité.

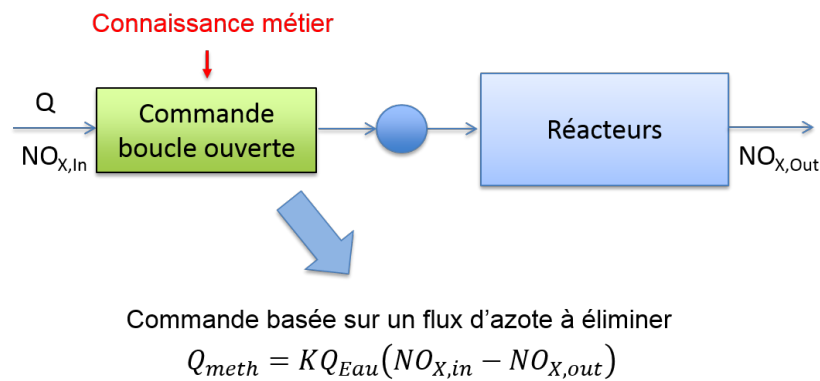

Commande basée sur un flux d'azote à éliminer
$$Q_{meth} = K Q_{Eau}(NO_{X,in} - NO_{X,out})$$

**Figure 5. Schéma de la boucle de régulation actuelle**

Tel que montré dans ROCHER et al. (2015), ce mode de régulation, basé sur une valeur de consigne en nitrates à respecter, ne permet pas de s'assurer de la stabilité du milieu réactionnel dans lequel baigne la biomasse épuratrice. Pour un coefficient d'exploitation fixe, le ratio C/N réel maintenu dans le réacteur biologique peut fluctuer dans des proportions importantes si le niveau de concentration de l'effluent alimentant l'ouvrage est fluctuant. Cette instabilité du milieu réactionnel perturbe le

fonctionnement de la biomasse épuratrice dont les besoins en carbone, azote et phosphore doivent, en toutes circonstances, être satisfaits pour assurer un traitement complet de l'azote nitrique. Ces perturbations conduisent à la production de nitrites.

## 3. PRISE EN COMPTE DU NITRITE RESIDUEL EN TEMPS REEL VIA LA COMMANDE SANS MODELE (CSM)

### 3.1 Description de la CSM

#### 3.1.1. Principe généraux

La commande sans modèle se place « en fin de traitement » et se base sur la concentration en nitrites mesurée en sortie de procédé. Elle se greffe à la méthode de contrôle classique précédemment décrite, en y apportant des corrections si besoin est (figure 6). Elle permet de fermer la boucle de contrôle par rapport au schéma actuel, ce qui habituellement fournit de meilleurs résultats. Bien qu'elle soit capable d'effectuer des corrections tant positives que négatives, la commande sans modèle a été restreinte à un apport positif uniquement dans cette étude. Plus directement, elle est inactive lorsque la concentration de nitrite en sortie de procédé passe sous la consigne, et redevient active lorsqu'au contraire la consigne est dépassée.

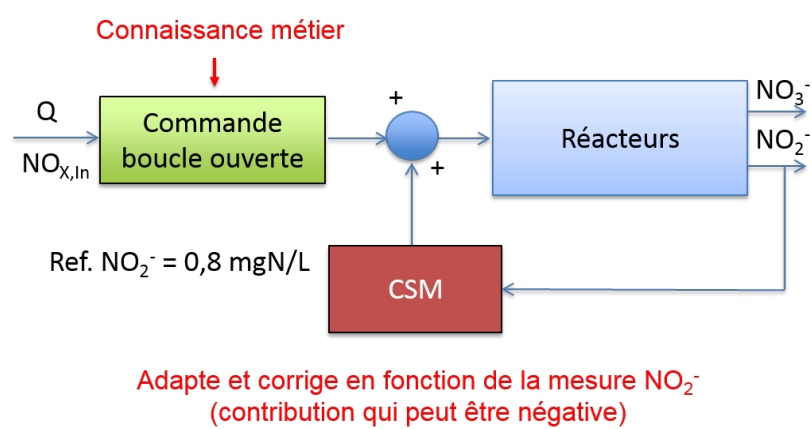

**Figure 6. Schéma de la boucle de régulation intégrant la commande sans modèle basée sur la mesure en nitrites résiduels en sortie d'ouvrage**

Une écrasante majorité des correcteurs utilisés en pratique sont de type PID (proportionnel-intégral-différentiel*).* L'utilisation fréquente de correcteurs PID est généralement motivée par le fait qu'ils ne nécessitent pas de ≪ bonne ≫ modélisation par des équations différentielles ou aux différences, trop souvent inextricable. Cependant, le réglage des gains et des correcteurs PID sont difficiles (ÅSTRÖM et MURRAY, 2008 ; FRANKLIN et al., 2015 ; de LARMINAT, 2012) et les performances de ces correcteurs se dégradent, parfois beaucoup, en présence de fortes non-stationnarités et non-linéarités.

Dans ce contexte, la « commande sans modèle » constitue une alternative intéressante. Cette commande repose sur des techniques nouvelles d'estimations en ligne (FLIESS et SIRA-RAMIREZ, 2008 ; SIRA-RAMIREZ et al., 2014). Elle a pour but de conserver les avantages des PID (absence de modélisation) tout en gommant leurs inconvénients. Des applications réussies dans les domaines les plus variés, accompagnées quelquefois de brevets, ont vu le jour, en France et à l'étranger. A titre d'exemple, on peut citer : les commandes de barrages hydroélectriques (JOIN et al., 2010), d'alliages à mémoire de forme (GEDOUIN et al., 2011), de manipulateurs flexibles (AGEE et al., 2015), de serres agricoles (LAFONT et al., 2015), et de certains convertisseurs (CAO et al., 2016). Mentionnons enfin la facilité et le coût faible de la mise-en-œuvre matérielle (JOIN et al., 2013 ; 2017). Le lecteur intéressé par la CSM trouvera en FLIESS et JOIN (2013) tous les développements nécessaires à une meilleure compréhension. La référence JOIN et al. (2017) contient des compléments techniques utiles quant à la dénitrification.

**3.1.2. Fondements mathématiques de la CSM**

**Modèle ultra-local.** On se restreint, pour simplifier les notations, à un système monovariable, d'entrée u (la variable manipulée, soit le débit de méthanol dans ce cas-ci) et de sortie y (la variable contrôlée, soit la concentration en nitrites en sortie de procédé dans ce cas-ci). On remplace le modèle mathématique global, qui est largement, pour ne pas dire entièrement, inconnu, par un modèle ≪ phénoménologique ≫, dit *ultra-local*, valable sur un court laps de temps,

$$y^{(\nu)} = F + \alpha u \qquad (1)$$

L'ordre de dérivation $v$, en général 1, choisi par l'opérateur, est étranger à l'ordre de dérivation maximum de $y$, inconnu, du système[1] ; le paramètre constant $\alpha$, fixé par l'opérateur afin que les valeurs numériques de $u$ et $y^{(v)}$ aient le même ordre de grandeur, n'a pas *a priori* de valeur précise ; $F$, qui contient toutes les informations « structurelles », dépend de toutes les autres variables du système, y compris des perturbations, et de leurs dérivées.

L'estimation en ligne réelle de la valeur numérique de $F$ permet de réactualiser (1) à chaque instant et confère à la commande qui en découle son caractère adaptatif. La comparaison suivante avec une procédure courante en informatique graphique devrait éclairer le lecteur. Pour reproduire une courbe arbitrairement compliquée sur un écran on n'utilise pas son équation mais l'approximation par de courts segments de droite juxtaposés. Le résultat satisfait l'œil en général. Ici, l'analogue de ce segment est (1).

**Correcteurs P intelligents.** L'utilisation du modèle développé en (1) permet de déterminer une valeur de u à injecter pour atteindre une valeur de y souhaitée. Ainsi, grâce au caractère prédictif de (1), l'évolution du nitrite en fonction de la dose de méthanol injectée peut être évaluée. Puisque ce modèle *ultra-local* est simple, il est facile de déterminer la dose de méthanol nécessaire à l'obtention d'une concentration résiduelle de $NO_2^-$ donnée. Le modèle n'étant toutefois jamais parfait, la valeur de débit de méthanol qui en découle est partiellement empreinte d'erreur.

Avec l'objectif d'imposer à $y$ de suivre une référence ou consigne $y*$, l'expression de $u$ issue de (1) s'écrit, pour **$v=1$** :

$$u = -\frac{F - \dot{y}^*}{\alpha} \qquad (2)$$

Afin de compenser une erreur éventuelle (liée à l'estimation de $F$, par exemple) un correcteur proportionnel est ajouté à l'action (2) :

$$u = -\frac{F - \dot{y}^* + K_P e}{\alpha} \qquad (3)$$

Où $e = y - y*$ est l'erreur de poursuite, $K_P > 0$ est un gain, choisi de sorte que $\dot{e} + K_P e = 0$ soit stable. La commande définie en (3) constitue alors un *proportionnel* rendu *intelligent* (ou iP) par le terme adaptatif *F*. La valeur de Kp doit être choisie pour obtenir un compromis satisfaisant entre rapidité de correction et stabilité du système en boucle fermée.

---

[1] Voir FLIESS et JOIN (2013) pour une discussion sur l'ordre $v$

La consigne *y*\* est déterminée par l'ingénieur de façon à obtenir la performance désirée, compte tenu des possibilités de l'appareillage. C'est l'aspect *prédictif*, ou *feedforward*. Quant au gain $K_P$ en (3), son choix est infiniment plus aisé que le réglage des PID, comme en conviendra tout utilisateur de cette technique.

Comme *F* en (1) contient tout ce qui est mal connu, le correcteur (3) est « robuste » par rapport à des phénomènes aussi complexes que les bruits de mesure, des perturbations ou le vieillissement (cette robustesse a été confirmée dans tous les exemples concrets). C'est un autre avantage considérable par rapport aux PID traditionnels.

Enfin, l'estimation de F est menée dans le domaine opérationnel[2] (YOSIDA, 1984), et est obtenue en suivant les percepts décrits dans FLIESS et SIRA-RAMIREZ (2008) et SIRA-RAMIREZ et al. (2014). Le lecteur pourra se référer à ces articles et, notamment à FLIESS et JOIN (2013) pour de plus amples détails sur cette question.

### 3.2. Evaluation de la capacité de la CSM à maitriser le flux de nitrites résiduels

**3.2.1. Description du modèle SimBio utilisé pour tester la CSM**

Les études menées pour cerner les processus d'apparition des nitrites lors de la dénitrification sur biofiltres (ROCHER et al., 2011a,b,c ; ROCHER et al, 2015) ont permis de générer les données nécessaires pour calibrer et valider un modèle de prédiction du fonctionnement des biofiltres dénitrifiant. Ce modèle (SimBio®, programme Mocopée), capable de simuler les performances des unités de de post-déntrification (BERNIER et al., 2014), a été utilisé comme plateforme pour tester l'efficacité de la commande sans modèle.

Il s'agit d'un modèle phénoménologique décrivant les processus globaux impliqués durant la biofiltration des eaux usées. Le modèle SimBio est construit sous l'environnement Matlab, avec la toolbox Simulink et à partir de sous-modèles disponibles dans la littérature scientifique. L'hydraulique

---

[2] La terminologie *transformation de Laplace* est aussi très courante (voir FRANKLIN et al., 2015 ; de LARMINAT, 2009 ; par exemple). Rappelons que *s*, dite parfois *variable de Laplace*, correspond à la dérivation par rapport au temps. Alors, $s^{-1}$ donne, dans le domaine temporel, l'intégrale

du biofiltre y est représentée à l'aide de six réacteurs complètement mélangés de volume égal placés en série, de manière à approximer un écoulement piston tout en maintenant les temps de simulation dans le domaine du raisonnable. La présence de média dans la colonne est considérée comme réduisant le volume disponible au passage du liquide dans les réacteurs par un facteur de porosité ε. Un biofilm est représenté sous formes de couches superposées sur le média à l'aide du modèle de SPENGEL et DZOMBAK (1992), adapté à la biofiltration. Le transport des polluants solubles vers, dans et hors du biofilm se fait par diffusion. Les polluants particulaires sont échangés entre la couche de surface du biofilm et le liquide par détachement du biofilm et filtration des particules (HORNER et al., 1986). L'impact des lavages sur l'enlèvement du biofilm est considéré simplement par une perte d'une fraction fixe de l'épaisseur du biofilm en surface, calculé de manière indépendante dans chaque réacteur.

Dans un contexte de post-dénitrification où on retrouve très peu de particules ou même de pollution carbonée soluble dans les eaux en entrée du procédé, le phénomène principal à modéliser consiste en la croissance des bactéries épuratrices dans le biofilm. La base du modèle de conversion biologique de la pollution est dans ce cas-ci l'Activated Sludge Model 1 (ASM1, HENZE et al., 1987). Ce modèle est fréquemment utilisé et représente la croissance des biomasses hétérotrophes et autotrophes, ainsi que la consommation et la production de substrats carbonés et azotés qui sont impliquées. Les réactions de nitrification et de dénitrification sont toutefois représentées dans l'ASM1 comme se déroulant en une seule étape, soit respectivement le passage du $NH_4^+$ en $NO_3^-$ directement, puis la transformation du $NO_3^-$ en $N_2$. Ceci fait en sorte que la version originale de ce modèle n'est pas en mesure de fournir des prédictions sur les concentrations en nitrites durant le traitement de l'azote. Afin de pallier ce manque, une modification est apportée au modèle de conversion biologique : la réaction de dénitrification est considérée par un processus se déroulant en deux étapes. Pour ce faire, les schémas du processus de dénitrification considérés dans le modèle ASMN (HIATT et GRADY, 2008) sont utilisés. Ce modèle est à la base conçu pour représenter la dénitrification en quatre étapes et ainsi pouvoir simuler les émissions de NO et $N_2O$ pouvant se produire. Une seule biomasse hétérotrophe est simulée, en considérant que seules certaines fractions sont en mesure de réaliser chacune des réactions impliquées. Seules les deux premières réactions de dénitrification du modèle ASMN ($NO_3^-$ vers $NO_2^-$ vers $N_2$) sont pour l'instant implantées dans le modèle SimBio. Celles-ci sont quelque peu adaptées du point de vue stœchiométrie pour compenser l'absence des deux dernières réactions.

La version du modèle SimBio utilisées dans le cadre de cette étude a précédemment été calibrée sur des mesures horaires de nitrites et nitrates récoltées sur une partie du procédé de post-dénitrification de la station d'épuration de Seine-Centre située à Colombes en banlieue parisienne. Plus de détails sont disponibles sur le modèle SimBio tant en nitrification tertiaire qu'en post-dénitrification respectivement dans ROCHER et al. (2014a,b) et dans BERNIER et al. (2014).

### 3.2.2. Stratégie d'évaluation de l'efficacité de la CSM

Afin d'en comparer l'efficacité, les deux stratégies de contrôle du nitrite précédemment décrites (figures 5 et 6) ont été implantées dans le modèle SimBio. Chaque méthode de contrôle a été simulée sur une période de dix jours, en utilisant un jeu de données synthétiques fortement représentatif des variations dans les charges de $NO_X$ observées en entrée de post-dénitrification à Seine-Centre. Les cinq premiers jours du modèle ont servi à acclimater le modèle aux conditions d'entrée imposées. Seuls les cinq derniers jours ont donc été considérés lors de l'évaluation des performances de chaque solution simulée. Afin de comparer l'efficacité des deux solutions de contrôle évaluées, la méthode de contrôle « actuelle » a été simulée en utilisant un jeu de paramètres de calcul de la dose de méthanol à injecter (rapport méthanol/$NO_X$) calibré. Cette calibration a visé à obtenir une concentration moyenne en nitrites en sortie de traitement la plus proche possible de la consigne indiquée à la commande sans modèle. La calibration des paramètres a dans ce cas été réalisée *a posteriori* sur le même jeu de données utilisé pour l'évaluation. Ce scénario illustre donc les performances optimales du contrôle actuel que l'on pourrait atteindre vis-à-vis des nitrites.

La stratégie de contrôle utilisant la commande sans modèle a quant à elle été couplée au mode de régulation actuel tel qu'illustré en figure 6. Les paramètres de la régulation actuelle ne sont dans ce cas-ci pas optimisés, l'amélioration des concentrations en nitrite dans l'effluent étant laissée entièrement à la commande sans modèle. Le coefficient d'exploitation est donc simplement fixé à sa valeur par défaut de 3.

### 3.2.2.1. Evaluation des résultats obtenus avec la CSM

La figure 7 illustre les données de concentrations en $NO_3^-$ imposées à l'entrée du modèle durant les cinq jours d'évaluation des performances de contrôle. La figure 8 illustre quant à elle les résultats des simulations pour les deux différents scénarios de contrôle évalués : le scénario actuel et le scénario commande sans modèle (CSM). Enfin, la figure 9 illustre pour les deux scénarios les répartitions en nitrites en sortie de traitement, en dose de méthanol utilisé et en nitrates en sortie de traitement, sous forme de boites à moustaches. Dans tous les cas, la consigne visée est de 0,8 mgN/L en nitrites.

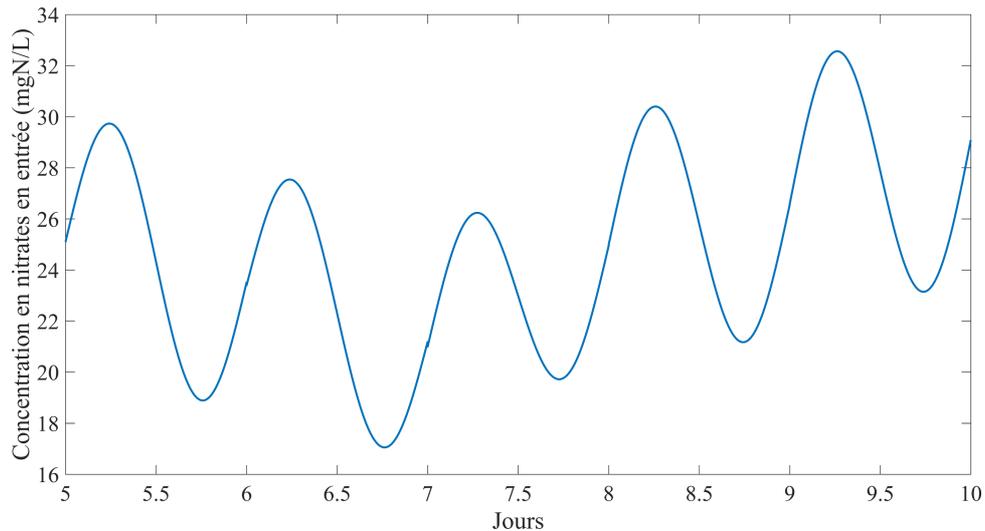

**Figure 7. Concentration en nitrates imposée en entrée du modèle SimBio lors des simulations des différents scénarios de contrôle**

Tel qu'illustré en figure 8, le contrôle actuel, une fois ses paramètres de calcul calibrés, parvient à osciller autour de la consigne visée (valeur moyenne simulée sur les nitrites de 0,84 mgN/L, médiane de 0,81 mgN/L). Ces oscillations sont par contre d'une ampleur relativement importante, atteignant un minimum et un maximum de respectivement 0,39 et 1,39 mgN/L de nitrite. Elles sont causées par le phénomène décrit précédemment : en utilisant des paramètres de calcul constants avec cette méthode de contrôle, le rapport C/N en entrée de traitement varie en fonction de la concentration en $NO_3^-$. Le modèle étant soumis à des variations intra et inter-journalières sur la teneur en nitrates (figure 7), ceci affecte la répartition $NO_3^-/NO_2^-$ dans l'eau traitée. Il est également important de rappeler que les paramètres de calcul utilisés pour obtenir ces résultats ont été optimisés *a posteriori* sur cette période de cinq jours pour viser l'atteinte de la consigne en moyenne autant que possible. Dans une situation réelle, ce degré d'optimisation est difficilement atteignable.

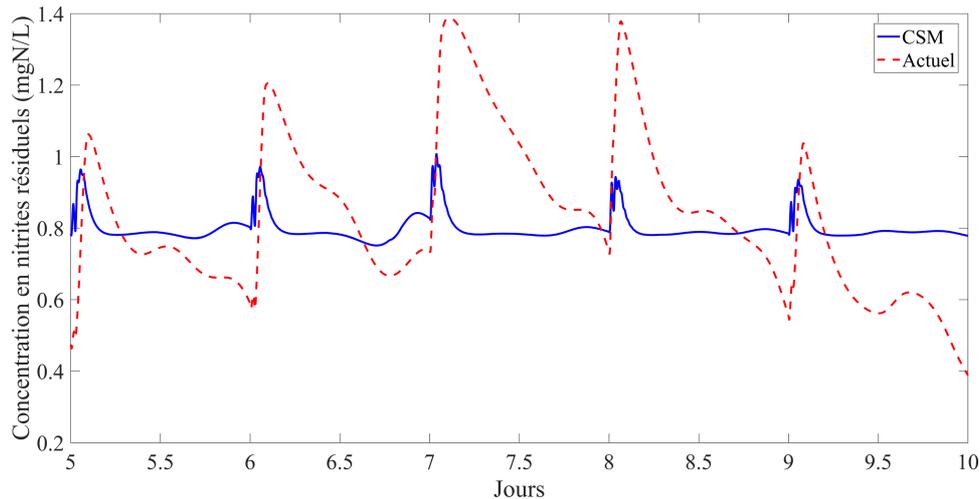

**Figure 8. Simulation des concentrations en nitrites résiduels avec le mode de régulation actuel (pointillé rouge) et le mode de régulation combinant la boucle métier et la commande sans modèle (trait bleu)**

La commande sans modèle parvient quant à elle à maintenir des concentrations en nitrites largement plus stables autour de la consigne. En effet, la valeur moyenne obtenue en sortie sur cette variable est de 0,80 mgN/L (médiane de 0,79 mgN/L), mais les valeurs minimale et maximale obtenues sont dans ce cas-ci de 0,75 et 1,00 mgN/L, respectivement. Seule une perturbation particulière est observable lors de chaque début de journée, où la teneur en nitrites en sortie augmente sur une courte période avant de revenir à la consigne. Celle-ci est due à la réalisation du lavage des biofiltres considérée se produire une fois par jour dans le modèle SimBio, en début de journée. Les lavages ont un effet potentiellement important sur les quantités de biomasse dénitrifiante dans le biofiltre et donc sur la production et l'élimination des nitrites et nitrates. Il semble que la commande sans modèle nécessite un certain moment avant d'arriver à se réadapter aux nouvelles conditions prévalant dans le réacteur. Cette période demeure toutefois relativement courte (~3 heures) et d'amplitude contenue, malgré l'effet important et soudain des lavages qui provoquent son apparition. Un effet similaire peut être observé sur les résultats de simulation obtenus avec le contrôle actuel. Il s'agit dans ce cas-ci d'un mélange de l'impact des lavages et de l'augmentation des concentrations de nitrate en entrée de procédé présente aux mêmes heures.

La boite à moustaches sur les nitrites en figure 9 confirme la nette amélioration sur la stabilité des $NO_2^-$ en sortie de traitement lors du passage à la méthode de contrôle actuelle au contrôle sans modèle. Les boites à moustache du centre (débit de méthanol injecté par le contrôle) et de droite (concentration en nitrates en sortie de traitement) de la figure 9 résument le comportement des autres variables importantes dans le contexte de la post-dénitrification. Dans le cas du méthanol injecté, les moyennes et médianes pour chaque méthode de contrôle sont relativement semblables (contrôle actuel : moyenne de 1181 kg/j et médiane de 1185 kg/j ; commande sans modèle : moyenne de 1181 kg/j et médiane de 1195 kg/j). Les quartiles 1 et 3 sont quant à eux légèrement plus élevés dans le cas de la commande sans modèle. Les gains de stabilité sur les nitrites apportés sont donc relativement peu impactant sur les coûts d'opération du procédé. Dans le cas du nitrate, la méthode de contrôle actuelle est en fait plus stable que la commande sans modèle. En effet, dans ce second cas, l'injection du méthanol est principalement dictée pour stabiliser les nitrites. Les variations de charge en nitrates en entrée de traitement se répercutent donc moins sur les nitrites qu'auparavant, mais plus sur les nitrates. Les concentrations en NOx ($NO_2^-$ + $NO_3^-$ dans ce cas-ci) simulées en sortie de traitement demeurent toutefois relativement similaires, quoique plus stables autour de la médiane pour la commande sans modèle que pour le contrôle actuel (non-montré). Ceci implique d'abord que le gain de stabilité sur les nitrites apporté par la commande sans modèle est dans l'absolu supérieur à la perte de stabilité sur les nitrates. Cela implique également qu'en dosant une quantité approximativement similaire de méthanol, mais à des moments plus appropriés, la commande sans modèle arrive à stabiliser à la fois les concentrations simulées en sortie de traitement en $NO_2^-$ et en NOx.

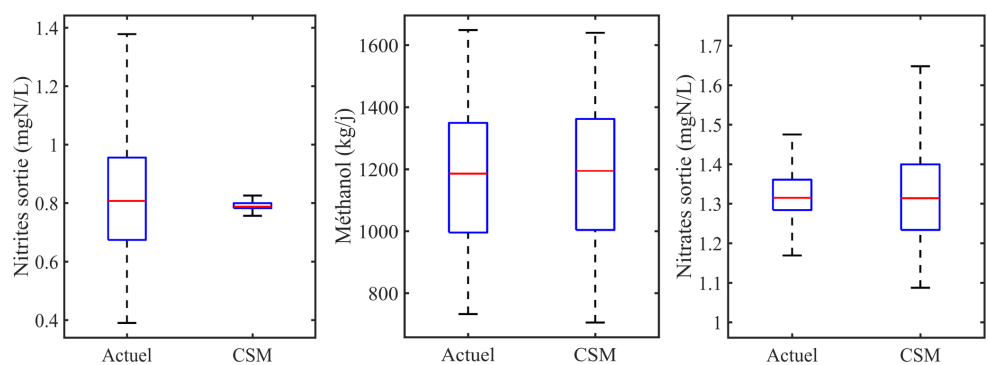

**Figure 9. Comparaison des performances simulées des deux modes de régulation au regard de la maîtrise des concentrations résiduelles en nitrites, de la consommation en méthanol et de la maîtrise des nitrates résiduels**

## 4. CONCLUSIONS

La modernisation des filières de traitement des eaux engagée ces 20 dernières années en agglomération parisienne a permis d'améliorer significativement la qualité de la Seine, notamment vis-à-vis de l'azote ammoniacal. Mais le développement des unités de post-dénitrification dans les stations d'épuration de l'agglomération parisienne a conjointement fait ré-émerger la problématique du nitrite dans les eaux de Seine. Ainsi, le contrôle de l'apparition des nitrites au cours de l'étape de post-dénitrification est devenu un enjeu technique majeur. Le SIAAP a engagé des études visant à mieux appréhender les mécanismes d'apparition du nitrite lors de la dénitrification des eaux usées et à étudier des évolutions techniques à mettre en œuvre sur les usines pour contrôler et limiter sa production (Programme MOCOPEE, www.mocopee.com).

Cet article montre que la modification des pratiques d'exploitation des unités de post-dénitrification permettrait d'accroître la maîtrise du nitrite en sortie de station d'épuration. Les modes de régulation classiques ne permettent pas de s'assurer de la stabilité du rapport C/N dans le réacteur biologique et conduisent à une production erratique et incontrôlée de nitrites. La possibilité d'ajouter une « commande sans modèle » à la commande classique a donc été testée à l'aide du modèle mathématique SimBio, modèle permettant de simuler le fonctionnement des unités de biofiltration. La commande sans modèle placée « en fin de traitement » et basée sur la concentration en nitrites mesurée en sortie de procédé, se greffe à la méthode de contrôle classique en y apportant des corrections seulement au besoin. Les résultats des simulations décrits dans ce papier montrent qu'une régulation des injections de méthanol basée sur la « commande sans modèle » permet de stabiliser et maitriser le nitrite dans le rejet, sans induire d'augmentation des quantités de méthanol injectées.

## REFERENCES BIBLIOGRAPHIQUES